# Design Paradigms of Intelligent Control Systems on a Chip

K. M. Deliparaschos and S. G. Tzafestas
Intelligent Automation Systems Research Group,
School of Electrical and Computer Engineering, NTUA,
Athens, Greece
kdelip@mail.ntua.gr, stzafesta@softlab.ntua.gr

*Abstract*— **This paper focuses on the Field Programmable Gate Array (FPGA) design and implementation of intelligent control system applications on a chip, specifically fuzzy logic and genetic algorithm processing units. Initially, an overview of the FPGA technology is presented, followed by design methodologies, development tools and the use of hardware description languages (HDL). Two FPGA design examples with the use of Hardware Description Languages (HDLs) of parameterized fuzzy logic controller cores are discussed. Thereinafter, a System-on-a-Chip (SoC) designed by the authors in previous work and realized on FPGA featuring a Digital Fuzzy Logic Controller (DFLC) and a soft processor core for the path tracking problem of mobile robots is discussed. Finally a Genetic Algorithm implementation (previously published by the authors) in FPGA chip for the Traveling Salesman Problem (TSP) is also discussed.**

I. INTRODUCTION

The rapid growth of Very Large Scale Integration (VLSI) technology and Electronic Design Automation (EDA) software tools in recent years has allowed for the development of high performance intelligent control systems for industrial and robotic applications. Modern EDA tools are used nowadays to create, simulate and verify the correct operation of a design idea for a complex system without the need of committing to hardware.

Field Programmable Gate Arrays (FPGAs) contain programmable logic components called "logic blocks", and a hierarchy of reconfigurable interconnects that allow the blocks to be "wired together". Using Hardware Description Languages (HDLs) such as VHDL and Verilog one can configure these logic blocks to perform complex logic structures. Many pre-written FPGA generic component cores exist that allow implementing processors, multipliers, video/audio converters, network protocols, etc. So an FPGA is extremely flexible when it comes to designing a complex system. FPGA technology is being used in several application fields such as telecommunications [1], signal [2] and image processing [3], medical equipment [4], automotive applications [5], robotics [6], [7], [8], [9], space landing crafts [10], just to name a few.

This paper aims to provide an overview of paradigms of the use of FPGAs to hardware realization of intelligent control systems that incorporate fuzzy logic theory, and genetic algorithms.

Section 2 attempts a brief introduction to the FPGA technology, HDLs and the development flow of a design targeting FPGA.

In Section 3, two examples of the FPGA design, implementation and functionality verification of a parameterized fuzzy logic processor core followed by a modified architecture (featuring a technique named by the authors as "Odd-Even" method) that achieves a significantly faster data processing rate are discussed [11], [12]. Next in Section 4, a scalable genetic algorithm processor core is presented and evaluated using the Traveling Salesman Problem (TSP) and several benchmarking functions [13], [14]. Finally in Section 5, a mobile robot for a path tracking application is controlled by a System on a Chip (SoC) combining a fuzzy tracker and a soft processor core [15], [16].

II. FPGA ARCHITECTURE AND HDLs

An FPGA is a semiconductor device that belongs to the family of programmable logic devices. The FPGA technology discussed in this paper is based on static memory or SRAM process technology while other technologies such as Flash (Flash-erase EPROM) exist.

The typical basic architecture consists of an array of Configurable Logic Blocks (CLBs) (combinatorial or/and sequential) and routing channels. Multiple I/O pads may fit into the height of one row or the width of one column in the array. Generally, all the routing channels have the same width (number of wires). An application circuit must be mapped into an FPGA with adequate resources.

A classic FPGA logic block consists of a 4-input Look-Up Table (LUT), a flip-flop, and a 2-to-1 multiplexer (to bypass the flip-flop if desired resulting in a registered or unregistered output) as shown in Fig. 1 below. The LUT is like a small RAM (RAM-based LUTs) and has typically 4 inputs, so can implement any logic gate with up to 4-inputs or used as a storage element. In recent years, manufacturers have started moving to 6-input LUTs in their high performance parts, claiming increased performance.

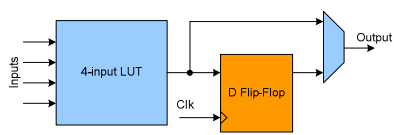

Figure 1. Configurable Logic Block (CLB).

Each logic block can be connected to other logic blocks through interconnect resources (wires/muxes placed around the logic blocks). Each block can do little, but with lots of them connected together, complex logic functions can be created. The interconnect wires also go to the boundary of the device where Input/Output Blocks (IOBs) are implemented and connected to the pins of the FPGAs.

In addition to general-purpose interconnect resources; FPGAs have fast dedicated lines in between neighboring logic blocks. The most common type of fast dedicated lines are carry chains. Carry chains allow creating arithmetic functions (like counters and adders) efficiently (low logic usage & high operating speed).

In addition to logic, all new FPGAs have dedicated blocks of static RAM distributed among and controlled by the logic elements. RAMs may be single port or dual port or even quad port. There are two types of internal RAMs in an FPGA namely, blockrams and distributed RAMs. The size of the RAM needed usually determines which type is used. The big RAM blocks, are blockrams which are dedicated areas in the FPGA. The small RAM blocks are either in smaller blockrams (Altera FPGAs), or in distributed RAM (Xilinx FPGAs). Distributed RAM allows using the FPGA logic-cells as tiny RAMs which provide a very flexible RAM distribution in an FPGA, but isn't efficient in term of area (a logic-cell can actually hold very little RAM). Altera prefers building different size blockrams around the device (more area efficient, but less flexible). Other features that have been observed more recently inside the FPGA architecture is the introduction of dedicated blocks such as DSP accelerators (hardwired multipliers with corresponding accumulators, high-speed clock management circuitry, and serial transceivers), embedded hard processor cores such as PowerPC [17] or ARM [18], and soft processor cores such as Nios [17] or Microblaze [18], [19]. Worth mentioned for control applications is the recent integration of an analog to digital converter in the Fusion mixed-signal FPGA from Actel [20]. However, this SoC trend does not replace the former generic architecture, but it can be seen as a complement to this original array.

FPGA pins are divided into 2 categories, dedicated pins and user pins (Input-Output or I/O pins). Dedicated pins are hard-coded to a specific function and are subdivided to power pins (ground or power), configuration pins (used to download to FPGA), and dedicated pins or clock pins. I/O pins can be programmed to be inputs, outputs, or bi-directional (tri-state buffers). Each I/O pin is connected to an IOB inside the FPGA. The IOBs are powered by the VCCO pins (I/O power pins). The FPGA VCCO pins (I/O power pins) are usually all connected to the same voltage. New FPGA generations feature user I/O banks in which I/Os are split into groups, each having its own VCCO voltage. This is useful for example if one part of the board works with 3.3V logic, and another with 2.5V.

FPGAs usually require two voltages to operate namely a core voltage and an IO voltage. Each voltage is provided through separate power pins. The internal core voltage or VCCINT, is used to power the logic gates and flip-flops inside the FPGA. The voltage can range from 5V for older FPGA generations, to 3.3V, 2.5V, 1.8V, 1.5V and even lower for the latest devices. The core voltage is fixed and is set by the FPGA model used. The I/O voltage or VCCO is used to power the IOBs of the FPGA and should match what the other devices connected to the FPGA expect. Specifically, FPGA devices allow VCCINT and VCCIO to be the same (i.e., the VCCINT and VCCIO pins could be connected together). But since FPGAs tend to use low-voltage cores and higher voltage IOs, the two voltages are usually different.

The latest FPGAs are produced using a 40-nm copper process [17] and their density can reach more than 680K logic elements or 13 million equivalent gates per chip with clock system frequencies approaching 600 MHz. Nevertheless, it is important to note that these specifications are only valid for a short while as technology evolves fast. The two major FPGA manufacturers are Xilinx [18], and Altera [17], while Lattice [21], Actel [20], Quicklogic [22] are smaller players.

To define the behavior of the FPGA, the user provides a Hardware Description Language (HDL) or a schematic design. The HDL form might be easier to work with when handling large structures because it's possible to just specify them numerically rather than having to draw every piece by hand. On the other hand, schematic entry can allow for easier visualisation of a design. Then, using an Electronic Design Automation (EDA) tool, a technology-mapped netlist is generated. The netlist can then be fitted to the actual FPGA architecture using a process called place-and-route, usually performed by the FPGA company's proprietary place-and-route software. The user will validate the map, place and route results via timing analysis, simulation, and other verification methodologies. Once the design and validation process is complete, the binary file generated (also using the FPGA company's proprietary software) is used to (re)configure the FPGA. This binary file is then transferred to the FPGA via a serial interface (JTAG) or to an external memory device like an EEPROM.

The most common HDLs are VHDL and Verilog, although in an attempt to reduce the complexity of designing in HDLs, there are moves to raise the abstraction level through the introduction of alternative languages such as SystemC (more of a system description language rather than hardware one), HandelC and SystemVerilog.

To simplify the design of complex systems in FPGAs, there exist libraries of predefined complex functions and circuits that have been tested and optimized to speed up the design process. These predefined circuits are called Intellectual Property (IP) cores, and are available from FPGA vendors and third-party IP suppliers.

In a typical design flow depicted in Fig. 2, one starts by modeling the behavior of the entire system (Electronic System Level – ESL) using a high-level language such as C, C++, or MATLAB prior to HDL coding of the RTL description of the design. During the ESL stage several test vector files could be extracted to be used during the verification stages.

Throughout a design process the design is simulated at multiple stages. Initially the HDL RTL description is simulated through a testbench running test vectors in order to verify the system's behavior and observe results. Following, the synthesis engine turns the RTL design into a design implementation in terms of logic gates (netlist file). At this point simulation may be run (post-synthesis simulation) to confirm that the synthesized circuit conforms to the initial behavior (also verify timing). Finally the design is laid out in the FPGA at which point propagation delays can be added and the simulation run again with these values back-annotated onto the netlist.

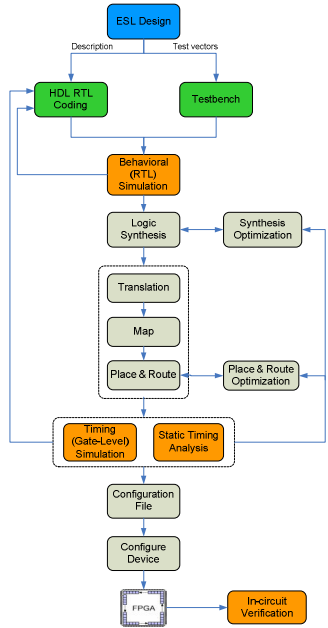

Figure 2. FPGA design flow.

### III. FPGA IMPLEMENTATION OF FUZZY CONTROLLERS

Fuzzy chips are distinguished into two classes depending on the design techniques employed: *digital* and *analog*. The first fuzzy chip was reported in 1986 at AT&T Bell Laboratories [23]. The digital approach originated from Togai and Watanabe's paper [23] and resulted in some interesting chips [24], [25]. Other digital architectures were reported in [26], [27]. Analog chip approaches begun with Yamakawa in [28], [29].

This section presents the design of a parameterized digital fuzzy logic controller (DFLC). By the term parameterized we mean that the DFLC facilitates scaling and can be configured for different number of inputs and outputs, number of triangular or trapezoidal fuzzy sets per input, number of singletons per output, antecedent method, divider type, and number of pipeline registers for the various components in the model. This parameterization enables the creation of a generic Fuzzy Processor (FC) core than can be used to produce fuzzy processors of different specifications without the need of redesigning the FC core from the beginning. The fuzzy logic processor architecture assumes overlap of maximum two fuzzy sets between adjacent fuzzy sets and requires $2^n$ clock cycles (input data processing rate, n is the number of inputs), since it processes one active rule per clock cycle. The architecture of the design allows one to achieve a core frequency speed of 100 MHz, while the input data can be sampled at a clock rate equal to $1/2^n$ of the core frequency speed (100 MHz), while processing only the active rules. To achieve this timing result the latency of the chip architecture involves 11 pipeline stages each one requiring 10 ns. The discussed DFLC is based on a simple algorithm similar to the Takagi-Sugeno (T-S) of zero-order type inference and weighted average defuzzification method [30], [31], and based on the chosen parameters employs four 12-bit inputs and one 12-bit output, with up to 7 trapezoidal or triangular shape membership functions per input with a degree of truth resolution of 8-bit and a rule base of up to 2401 rules.

It is well known that T-S fuzzy controllers can provide an effective representation of complex nonlinear systems in terms of fuzzy sets and fuzzy reasoning. The T-S method is considered a quite simple method, leads to fast calculations and is relatively easy to apply. Moreover a fuzzy controller based on the T-S method provides a good trade-off between the hardware simplicity and the control efficiency. In the T-S inference rule, the conclusion is expressed in the form of linear functions. Rules have the following form:

Rule $R_i$: IF $x_1$ IS $A^1_i$ AND … AND $x_k$ IS $A^k_i$

THEN $y_i = c^0_i + c^1_i x_1 + … + c^k_i x_k$

where, $x_1,…x_k$ represent the input variables, $A^1_i,…A^k_i$ represent the input membership functions, $y_i$ represent the output variable, and $c^0_i,…c^k_i$ are all constants. A zero-order model arises (simplified T-S functional reasoning), if we only use the constant $c^0_i$ at the output, hence $y_i=c^0_i$ (singleton outputs). In the discussed T-S model, inference with several rules proceeds as usual, with a firing strength associated with each rule, but each output is linearly dependent on the inputs. The output from each rule is a moving singleton, and the deffuzified output is the weighted average of the contribution of each rule, as shown in the following equation (Eq.1):

$$y = \frac{\sum_{i=1}^{m} w^i y^i}{\sum_{i=1}^{m} w^i} \qquad (1)$$

where $w^i$ is the weight contribution of the left part of the $i^{th}$ rule and for AND method connection is given by Eq.2:

$$w^i = \prod_{k=1}^{n} \mu_{A_k^i}(x_k) \text{ or } w^i = \min_{k=1}^{n} \mu_{A_k^i}(x_k) \quad (2)$$

A block diagram illustrating the T-S mechanism is depicted in Fig. 3.

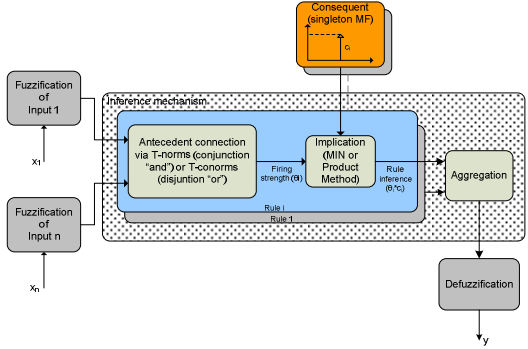

Figure 3.  Takagi-Sugeno (T-S) mechanism.

The Table 1 below summarizes the characteristics of the parameterized DFLC based on the chosen parameters (VHDL package file).

TABLE I.  DFLC SOFT CORE CHARACTERISTICS

| Fuzzy Inference System (FIS) type | Takagi-Sugeno zero-order type |
|---|---|
| Inputs | 4 |
| Input resolution | 12-bit |
| Outputs | 1 |
| Output resolution | 12-bit |
| Antecedent Membership Functions (MF's) | 7 Triangular or Trapezoidal shaped per fuzzy set |
| Antecedent MF degree of truth ($\alpha$ value) resolution | 8-bit |
| Consequent MF's | 2401 Singleton type |
| Consequent MF resolution | 8-bit |
| Maximum number of fuzzy inference rules | 2401 ( $no.of\ MFs^{no.of\ IPs}$ ) |
| AND method | MIN (Gödel minimum t-norm) |
| Implication method | PROD (Larsen product t-norm) |
| MF overlapping degree | 2 |
| Defuzzification method | Weighted average |

The DFLC architecture [11] (shown in Fig. 4) is composed of a number of blocks coded in a parameterized style to result in a fully parametric core. The design architecture is broken in three major hierarchical blocks, namely: "Fuzzification", "Inference" and "Defuzzification". The "CPR" blocks represent the number of pipeline stages for each component (component pipeline registers), the "PSR" blocks indicate the path synchronization registers, while the "U" blocks represent the different components of the DFLC. The component named after DCM (U0) is the Digital Clock Manager and is one of four components available in the specific FPGA library chosen (Spartan-3 1500-4FG676) [18]. The latter accepts the FPGA board (Memec 3SMB1500) clock (clk) of 75 MHz and is configured to generate a divided clock signal (clkdv) of 75 MHz/12=6.25 MHz and a multiplied clock signal (clkfx) which is 6.25·16=100 MHz for the DFLC core operation. It is worth mentioning here that the above frequency values can vary by adjusting the DCM generic parameters. A control logic component (control_logic_p, U2) is implemented on the top structural entity (binds FC core, DCM, R1, R2 and control logic) of the design that provides two signals named "ready_in" and "ready_out", used to provide input and output data-ready signals for handshaking with external devices.

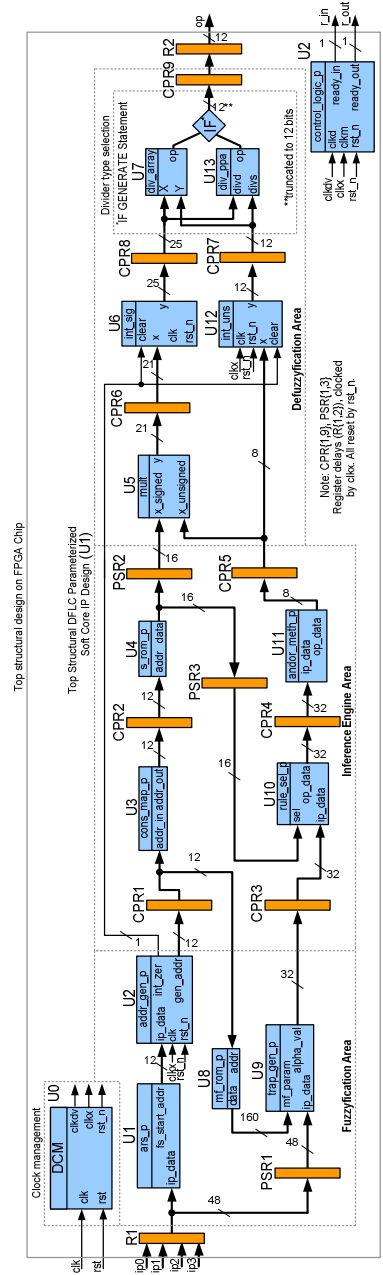

Figure 4.  Parameterized DFLC soft core architecture.

Processing all rule combinations for each new data input set would greatly reduce the overall processing rate, for this reason we have chosen to process only the active rules, i.e. those rules that give a non null contribution to the final result. Given the fact that the overlapping between adjacent fuzzy sets is of order 2, the described processing bottleneck is overcome by using an *active rule selection* block (*ars_p*) to calculate the fuzzy set region in which the input data corresponds to. As a consequent, for the 4-input DFLC with 7 membership functions per input, only 16 active rules need to be processed instead of all 2401 rule combinations in the rule base,.

The simulation snapshot in Fig. 5 depicts the data flow for all the signals specified in Fig. 4. A newly arrived input data set is clocked on the rising edge of the *clkdv_out* clock, with a fraction of 1/16 the frequency of the *clkfx_out* clock, providing the necessary time (16 algorithmic clock cycles) for the Address Generator (pipeline stage, cpr1) to generate the signals corresponding to all active rules ($2^{nd}$-$17^{th}$ clock cycles). Here, we remind that by using the *ars_p* block we effectively identify and process 16 active rules per clock cycle instead of 2401.

The total data processing time starting from the time a new data set is clocked at the inputs until it produces valid results at the output requires a processing time of 270 ns which is analyzed in 16 algorithmic clock cycles (each active rule is processed in one clock cycle) and 11 clock cycles due to pipelining. This effectively characterizes the input data processing rate of the system (a new valid data set is given at the output every 16 clocks or 160 ns), while the DFLC core operates at an internal clock frequency operation of 100 MHz or 10 ns period. Along with the proposed DFLC architecture, a modified model with LUT based Membership Function (MF) generator blocks, instead of arithmetic based, has been implemented which is not analyzed here. The latter DFLC model achieves a better timing result with the same levels of pipelining, but with significant increase in FPGA area utilization compared to the discussed model. It is obvious that since the MF's are ROM based any type and shape could be implemented.

To further improve the data processing rate of afore mentioned DFLC a method named "Odd-Even" by the authors [12] has been used to effectively separate the fuzzy sets to odd and even regions. The modified DFLC featuring the discussed method features two 8-bit inputs and a 12-bit output, with up to 7 trapezoidal or triangular shape membership functions per input and a rule base of up to 49 rules (see Fig. 6).

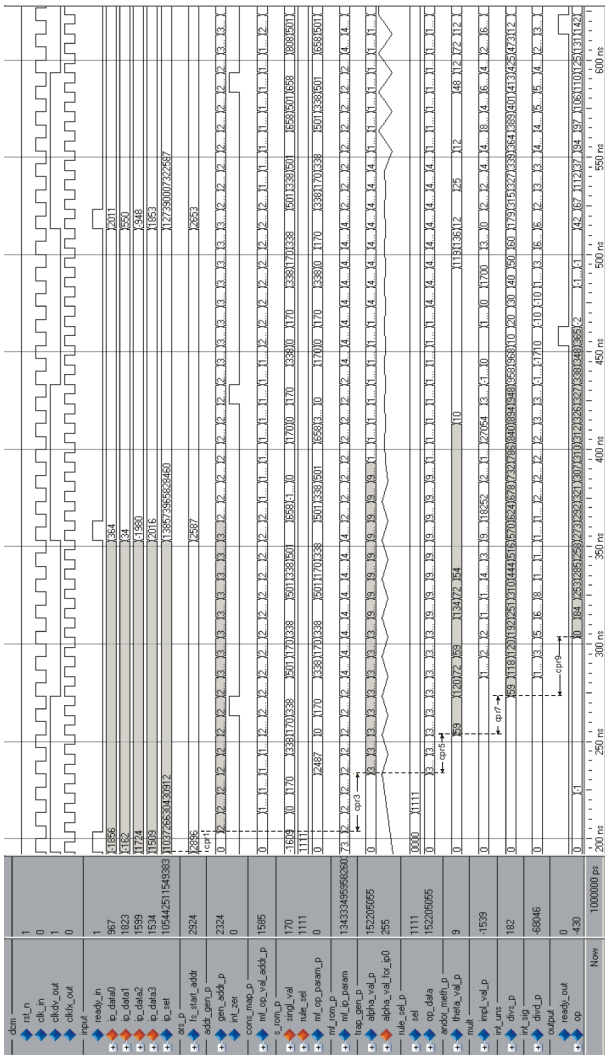

Figure 5.  Pipelined data flow

Figure 6.  DFLC architecture with "Odd-Even" method.

Using the "Odd-Even" method, we manage to process for the same model case scenario, two active rules per clock cycle, thus increasing significantly the input data processing rate of the system to $2^n/2$, where $n$ is the number of inputs. The architecture of the design achieves a core frequency speed of 200 MHz, allowing the input data to be sampled at a clock rate equal to half of the core frequency speed (100 MHz). The total processing time of the chip architecture involves 13 pipeline stages each one requiring 5 ns.

It's worth mentioning here that all the VHDL codes for the DFLC models are fully parameterized allowing us to generate and test DFLC models with different specification scenarios.

## IV. FPGA IMPLEMENTATION OF GENETIC ALGORITHMS

Genetic Algorithms (GAs), initially developed by Holland [32], are based on the notion of population individuals (genes/chromosomes), to which genetic operations as mutation, crossover and elitism are applied. GAs obey Darwin's natural selection law i.e., the survival of the fittest. GAs have been successfully applied to several hard optimization problems, due to their endogenous flexibility and freedom in finding the optimal solution of the problem [33], [34].

However, the most serious drawbacks of software-implemented GAs are both the vast time and system resources consumption. Keeping that in mind, a multitude of hardware-implemented GAs have been evolved mainly during the last decade, exploiting the rapid evolution in the field of the FPGAs technology and achieving impressive time-speedups.

This section deploys the design and hardware implementation of a parameterized GA IP core on an FPGA chip [13], [14]. The genetic operators applied to the genes of the population are crossover, mutation and elitism, whose employed method is parametrically selected. The designed selection algorithm is the "Roulette Wheel Selection Algorithm". A software implementation of the designed GA using the MATLAB platform has also been developed to produce input and output test vectors for the performance evaluation of the hardware implemented GA using several benchmark functions. Finally, after adapting the proposed hardware implemented GA to the Travelling Salesman Problem (TSP), a successful solution to it has been found. The evaluation of the algorithm was performed using the TSP and several benchmark functions.

A high level view of the architectural structure of the presented GA is shown in Fig. 7. The system is composed of six basic modules i.e. *control module*, *fitness evaluation module*, *selection module*, *crossover module*, *mutation module* and *observer module*. The control module implements a Mealy state machine, which feeds all other modules with the necessary control signals guaranteeing their synchronized execution. The selection module implements the Roulette Wheel Selection Algorithm [33], [34] picking the genes of the current population (*parents*), which will be genetically processed in order to create the individuals of the new population (*offsprings/children*).

Following, the crossover and mutation modules perform the corresponding genetic operations on the selected parents of the current generation. Thereafter, the fitness evaluation module not only computes the fitness of each offspring produced by the previous mentioned modules but also applies elitism to them, creating the elite genes for the next generation. The task of the observer module is to determine if the stopping criteria of the GA i.e., maximum number of generations, fitness limit, have been met so as to decide the continuation or not of the algorithm.

Four random number generators (RNG) are also used to produce both the initial random generation and the necessary random numbers. Additionally, one RAM is needed for the storage of the current gene population (RAM 1) and another one for the storage of the selected parents (RAM 2) of each generation.

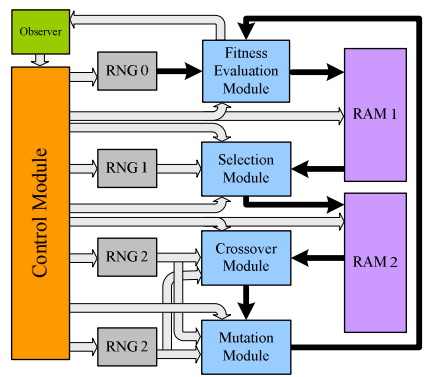

Figure 7. Architectural structure of the GA (high level view).

The parameters of the GA are summarized in Table 2.

TABLE II. CHARACTERISTICS OF THE GA IMPLEMENTATION

| Parameter name | Description | Possible Value |
|---|---|---|
| genom_lngt | Chromosome length in bits | 16 |
| score_sz | Fitness value bit resolution | 16 |
| pop_sz | Population size | 32 |
| scaling_factor_res | Bit resolution of the random number used in RWS algorithm | 4 |
| elite | Number of elite children | 2 |
| mr | Mutation rate | 80 |

The GA architecture (Fig. 8) is broken into separate blocks each one of which performs a particular task, coordinated by the control block. Moreover, the blocks communicate with the control module notifying their state i.e., ready out signals. Signal and data buses are noted on the block diagram with tenuous and bold lines respectively.

A performance evaluation of the GA using the TSP has been performed by comparing the time needed to find the optimal solution using the software version vs. the hardware implementation of. Table 3 summarizes the results for eight cities, 60 generations and 32 individuals where an impressive speedup ratio of 11,035 can be observed. Moreover, the algorithm was tested using the benchmark *burma14* derived from the TSPLIB [35] (analytical results presented in [13]).

TABLE III. SOFTWARE VS. HARDWARE IMPLEMENTED GA

| GA version | Time (ms) |
|---|---|
| Hardware (clk =10.8 ns) | 1.702 |
| Software (Pentium 4 3.2Ghz 1Gb RAM) | 18,783 |

The performance of a GA can be evaluated by several benchmark functions found in literature [36], [37], i.e. its ability to reach the optimum of an objective function. The GA core discussed here has been tested using the before mentioned functions. The results are analyzed in [13].

The GA IP core is fully parameterized in terms of the number of population individuals (*pop_sz*) and their resolution in bits (*genom_lngt*), resolution in bits of the fitness (*score_sz*), number of elite genes in each generation (*elite*), method used for crossover (*cross_method*) and mutation (*mut_method*), number of maximum generations (*max_gen*), mutation probability (*mr*) and its resolution in bits (*mut_res*), as well as the resolution in bits of the scaling factor *r* used by the RWS algorithm. The core parameterization allows the adaptation of the GA to any problem specifications without any further change to the developed VHDL code. The hardware implemented GA operates at a clock rate of 92 MHz (10.8 ns) and achieves a remarkable speedup when compared to its software version implemented in MATLAB. Furthermore, the FPGA resources regarding area and RAM requirements are kept small according to the place and route report [13], [14].

The presented IP core design when compared to other GAs hardware implementations [38], [39], [40], [41], operates at a clock frequency up to five times faster and implements more than one crossover and mutation methods, which can be changed during its execution. In addition, the core utilizes more parameters and is evaluated not only by using benchmarking functions but also by solving the NP–complete TSP.

V. SoC FOR ROBOT PATH TRACKING

A SoC implementation for robot path tracking [15], [16] on a differential-drive Pioneer 3-DX8 mobile robot is presented in this section. The SoC mainly consists of a DFLC core implementing the fuzzy tracking algorithm and a Xilinx Microblaze soft processor core acting as the top level flow controller.

The FPGA board hosting the SoC was attached to an actual differential-drive Pioneer 3-DX8 robot that has been used in running experiments of the tracking scheme. The developed SoC attains a core frequency speed of 71 MHz. The input data to the DFLC IP can be sampled at a clock rate equal to $1/2^n$ of the core frequency speed, and processing is performed accounting for only the active rules. To achieve this timing result the latency of the chip architecture involves 9 pipeline stages each one requiring 14.085 ns.

The fuzzy tracking algorithm used, is based on a previous fuzzy path tracker developed by the authors [42]. The fuzzy logic (FL) tracker has undergone some alterations due to the hardware restrictions posed by the DFLC soft IP core. While the original fuzzy logic controller (FLC) was Mamdani-based

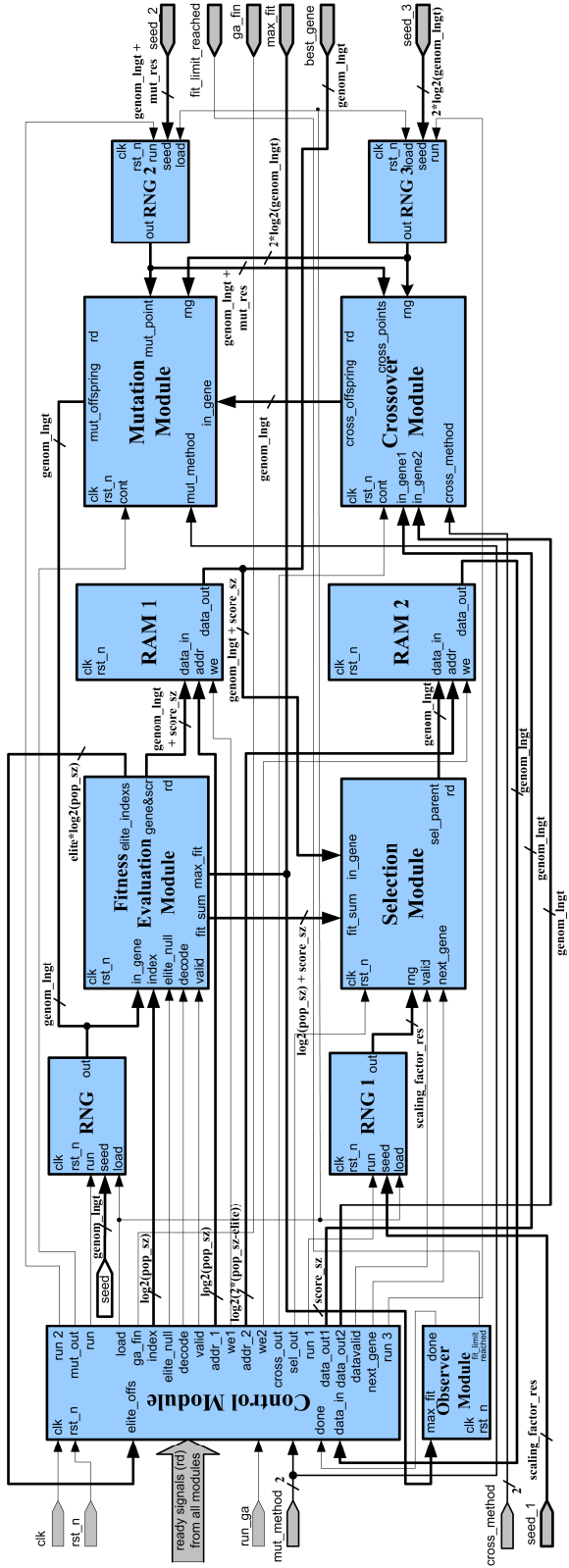

Figure 8. GA architecture

[43]with Gaussian membership functions, the one deployed here is a Takagi-Sugeno zero-order type FLC with triangular membership functions and an overlap of two between adjacent membership functions. The parameterized DFLC IP core discussed in Section III is utilized here and according to the specifications of the fuzzy tracker model, its parameters were appropriately set to form the following characteristics: two 12-bit inputs, one 12-bit output, and 9 triangular shape membership functions per input with a degree of truth resolution of 8 bits and a rule base of up to 81 rules [15], [16]. Further, the "spatial window" technique that was also introduced in [42] has been incorporated in the tracking scheme.

The system consists of four modules tied together. An overview can be seen in Fig. 9 (a), and an actual depiction in Fig. 9 (b).

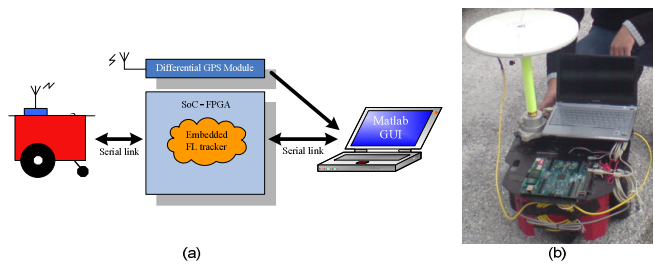

Figure 9. Overview of the system (b) Actual depiction of the system during an experiment. The FPGA, laptop and GPS antenna are clearly visible while the GPS receiver is under the laptop

The SoC implements the autonomous control logic of the P3 robot. It receives odometry information from the robot and issues steering commands outputted by the FL tracker. Several other tasks realized by the SoC besides the steering control include: (i) decoding the information packets sent by the robot which include the pose estimation done by the robot, (ii) the status of the motors, and (iii) encoding the steering commands in a data frame that is accepted by the robot. So, in other words, the SoC implements a codec for the I/O communication with the P3 robot. Additionally, it also relays some critical information to a MATLAB monitoring program that has been developed. The top-level program that attends to all these tasks and also handles synchronization and timing requirements is written in C and executed by the Microblaze soft processor core.

The ActivMedia P3-DX8 robot [15], [16] is the platform where the SoC was tested on. The robot employs a 1 mm resolution for the position estimation and 1° angle resolution for the heading. The kinematics of the robot are emulated to a bounded curvature steering vehicle and not that of a differential drive one i.e., there is an imposed constraint on the maximum curvature it can turn with. This constraint is a result of the fuzzy tracking algorithm being intended for the Dubin's Car model [44] where there is a minimum turning radius constraint on the robot and only forward motion.

The robot connects to the FPGA board through a serial cable to send and receive framed data. ActivMedia uses its own data framing protocol handled by the robot's microcontroller. The data sent from the robot are named *Server Information Packets* (SIP packets) while the received data are named *Command Packets*. More information on the data framing protocol can be found in the robot's manual [45]. Experiments clearly showed that even though the FL tracker performs well, its actual performance is severely degraded by the odometry's accumulation of errors over time. Several calibration tests were carried out in order improve odometry localization but as it was expected, position estimation through odometry proved ineffective.

A high level view of the proposed SoC architecture is illustrated in Fig. 10.

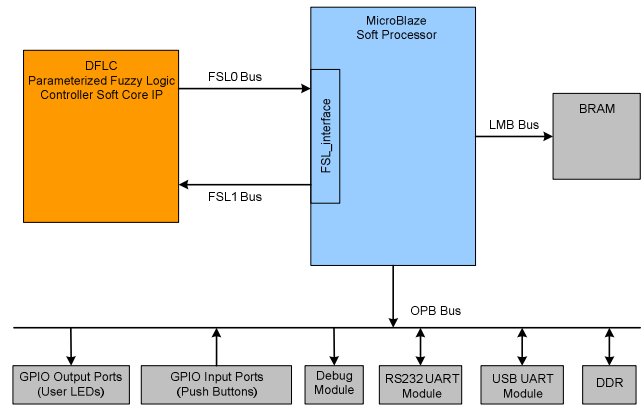

Figure 10. SoC high-level hardware system view.

The architecture of the SoC consists mainly of the DFLC that communicates with the Microblaze processor through the Fast Simplex Bus (FSL), the utilized block RAMs (BRAM) through the LMB bus, and other peripherals such as the general purpose input/output ports (GPIO), and UART modules via the OPB bus. The DFLC incorporates the fuzzy tracking algorithm, whereas the Microblaze processor mainly executes the C code for the flow control.

The U_fpga_fc component is embedded in the flc_ip top structural entity wrapper which is compliant with the FSL standard and provides all the necessary peripheral logic to the DFLC soft core IP in order to send/receive data to/from the FSL bus. The flc_ip wrapper architecture is shown in Fig. 11.

A MATLAB application was developed for monitoring and initialization purposes. The MATLAB application displays useful information about the robot's pose and speed, as estimated by the robot's odometry, as well as some other data used for the path tracking control. Furthermore it calculates the robot's position relative to the world and the local coordinate systems.

The application communicates with the FPGA board through a bridged USB connection. It receives and analyzes data relayed by the SoC, mainly the SIP packets that the robot sends. The program decodes the SIP packets and extracts odometry information. It also incorporates the same routine used in the SoC for catching and fixing encoder overflows.

Moreover, the MATLAB application incorporates the main world frame and carries out transformations from local to global coordinates. The GUI depicts the world map in global coordinates, as illustrated in Fig. 12.

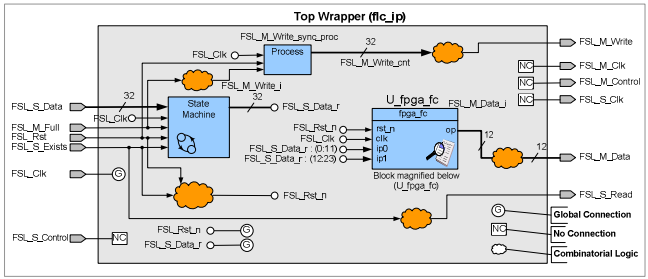

Figure 11. Top wrapper.

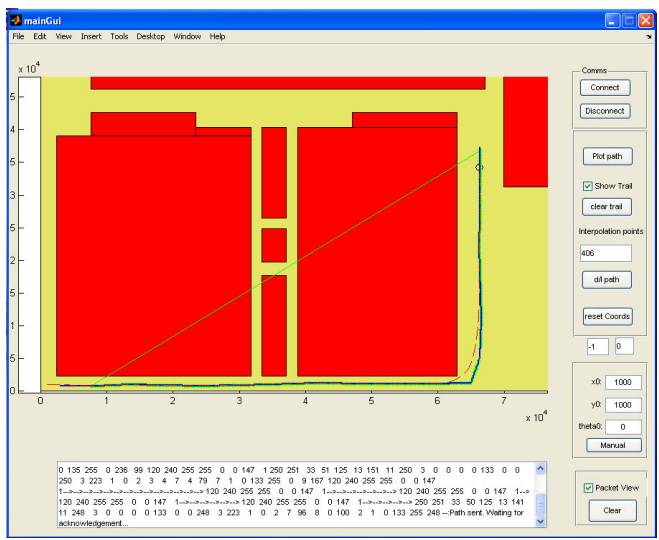

Figure 12. Snapshot of the GUI after an experiment. The solid line represents the desired path while the dashed line the actual path. The map illustrates part of the 2nd floor of the Electrical & Computer Engineering faculty of NTUA. All units are in millimetres.

An additional important function of the MATLAB application is to provide a path for the robot to track. Since the current application deals only with the path tracking task and not path planning routine, the path is drawn in the application's GUI by hand as a sequence of points encoded properly and downloaded to the SoC. At that point in time, the SoC begins the tracking control. The program uses a linear interpolation scheme to produce all the data samples of the path under a fixed sampling spacing, i.e., the distance between two sample points on the path is constant. The application allows choosing the number of interpolation points. The aforementioned interpolation routine was chosen after field observations on different interpolation schemes such as polynomial, cubic and linear and produced the best results.

The MATLAB GUI depicts the pose of the robot in real time along with other information sent by the SoC. In particular, when the *spatial window* is of order one, i.e., when only the closest point is considered, the SoC sends the two calculated controller inputs.

Next, the results of two experiments of the system are discussed. The experiments took place inside the NTUA campus. The goal was to assess the overall efficiency of the system and particularly of the fuzzy tracker. The experiments include tracking two prescribed paths. The first one is a straight line path while the second one is an S-shaped path. In order to log the actual position of the robot during the runs, a DGPS antenna and receiver was mounted onto it. The DGPS system used is the Trimble 4700 GPS receiver. The GPS was set to Kinematic Survey mode where the path is solved in post-processing, not real time. In this mode the horizontal precision is ±1cm+1ppm for a baseline under 10Km. The occupation is 1 second i.e., a positional sample is calculated each second. The results of the two experiments for the straight and the S-shape runs are shown in Fig. 13 (a) and (b) respectively.

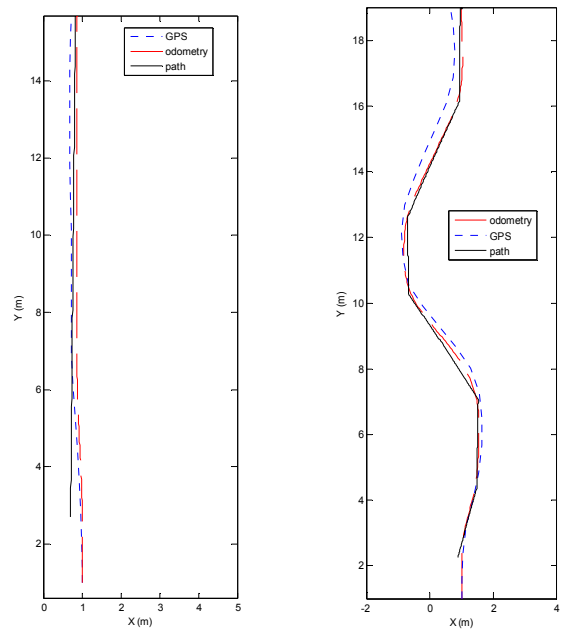

Figure 13. (a) The straight run experiment with the reference path (solid), the odometry position estimation (dashed) and the DGPS estimation (dotted). (b) The S-shaped path experiment with the reference path (solid), the odometry position estimation (dashed) and the DGPS estimation (dotted).

In the straight run experiment the robot was set to follow a 25m straight path while its initial position was not on the path. The depiction in Fig. 13 (a) does not present the entire run, but rather the segment where the GPS solution is of the highest quality (quality factor Q=1) since in order to assess the path tracker's performance we need a high precision position estimation. This must not be confused with the position estimation module that the tracker uses, which in this case is derived from odometry data. The GPS is used in order to see the actual position of the robot, thus a degraded GPS solution is useless and positional data of a Q factor greater than 1 (with 1 being the best and 6 the worst) have been discarded.

The second run in Fig. 13(b) presents the tracking of an S-shaped path. The same conditions regarding the GPS data also apply to this run. The S-shaped path has a length of approximately 25m. All GPS data with Q>1 have been discarded. It is evident from both experiments that the path tracker performs well. It should also be noted that the S-shaped path is not actually a *feasible* reference path since the curvature derivative is discontinuous at the polygons vertices. However if the discontinuity is small, the robot is expected to provide an accurate tracking. Furthermore the odometry position estimation is very close to the path. This means that if higher precision position estimation is used with the path tracker, such as a Real-Time Kinematic DGPS data feed that provides positional data to the path tracker in real-time, the tracker will perform even better. This is part of the future work of the authors. Moreover, the FPGA can easily incorporate data from other sensors and provide additional output.

## VI. Conclusion

FPGA technology, HDLs and EDA tools in recent years has allowed for the development of high performance intelligent control systems for industrial and robotic applications. Modern EDA software tools are used nowadays by the designers to create, simulate and verify the correct operation of a model of a complex system without the need of committing to hardware.

In this work several intelligent control system applications implemented on FPGA chips were presented. Three parameterized reusable FPGA cores, among them two fuzzy logic processors [11], [12] and a genetic algorithm processor unit [13], [14], previously developed by the authors, were discussed. Furthermore, a SoC for a path tracking task application on a differential-drive Pioneer 3-DX8 mobile robot [15], [16] was presented, that successfully utilizes the previously developed parametric DFLC core. The DFLC interfaces with a soft processor core and other secondary cores. The scalability of the fuzzy logic processor core [11] easily allowed adapting it to the fuzzy tracker model [42] without the need of recoding the core.